# Mining Association Rules in Various Computing Environments: A Survey

**Sudhakar Singh**
*Department of Computer Science, Institute of Science,
Banaras Hindu University, Varanasi 221005, India.*

**Pankaj Singh**[*]
*Faculty of Education, Banaras Hindu University,
Varanasi 221005, India,
[*]psingh.edu@bhu.ac.in*

**Rakhi Garg**
*Mahila Maha Vidyalaya, Banaras Hindu University,
Varanasi 221005, India,*

**P. K. Mishra**
*Department of Computer Science, Institute of Science,
Banaras Hindu University, Varanasi 221005, India.*

**Abstract**
Association Rule Mining (ARM) is one of the well know and most researched technique of data mining. There are so many ARM algorithms have been designed that their counting is a large number. In this paper we have surveyed the various ARM algorithms in four computing environments. The considered computing environments are sequential computing, parallel and distributed computing, grid computing and cloud computing. With the emergence of new computing paradigm, ARM algorithms have been designed by many researchers to improve the efficiency by utilizing the new paradigm. This paper represents the journey of ARM algorithms started from sequential algorithms, and through parallel and distributed, and grid based algorithms to the current state-of-the-art, along with the motives for adopting new machinery.

**Keywords:** association rule mining, frequent itemset mining, parallel computing, distributed computing, grid computing, cloud computing, MapReduce.

## Introduction

Association rule mining (ARM) [1] is one of the most famous technique of data mining, have received a wide attention in many areas like marketing, advertising, scientific and social network analysis and many others. ARM is based on Frequent Pattern Analysis which finds interesting correlation or association among attributes/items in large datasets. ARM technique has been first introduced by Agrawal et al. in 1993 [1] and they have described the formal model of association rule mining problem as follows.

Let $I = \{i_1, i_2, i_3, i_4, i_5, ..., i_m\}$ be a finite set of items. An *itemset* is defined as a collection of zero or more items of $I$ while *k-itemset* contains $k$ items of $I$. Let $D = \{T_1, T_2, T_3, ..., T_n\}$ be a finite set of transactions called datasets. Each transaction $T_i$ in database $D$ is an itemset such that $T_i \subseteq I$. The number of items present in a transaction is known as transaction width of that transaction. Let $X$ be a subset of set of items $I$, a transaction $T_i$ contains $X$ if $X \subseteq T_i$. The support count of $X$ is a number denoted by *supcount (X)*, defined as *supcount (X) = |{$T_i$ | $X \subseteq T_i$, $T_i \in D$}|* and the support of an itemset $X$ is the number *supp (X) = supcount (X) / n,* where *n* is total number of transaction in database.

An association rule is a conditional implication of the form $X \rightarrow Y$ where $X, Y \subset I$ are itemsets and $X \cap Y = \emptyset$. The strength of the rule is measured in terms of support and confidence denoted by *supp (X $\rightarrow$ Y)* and *conf (X $\rightarrow$ Y)* respectively and defined as *supp (X $\rightarrow$ Y) = supcount (X $\cup$ Y) / n* and *conf (X $\rightarrow$ Y) = supcount (X $\cup$ Y) / supcount (X)*. One has to find all the rules such they have *support $\geq$ min_sup* and *confidence $\geq$ min_conf*, where *min_sup* and *min_conf* are user defined threshold value for minimum support and minimum confidence [2-3]. Confidence of rule $X \rightarrow Y$ or $Y \rightarrow X$ can be easily obtained from the support count of *X, Y* and $X \cup Y$ if they are already computed. So mining association rules consists of two integral processes, the first one is frequent itemsets mining and the second one is mining strong association rules from frequent itemsets. The Frequent Itemsets Mining (FIM) is more computational intensive than the rules generation. So majority of the literatures are central to designing fast and scalable algorithms for frequent itemsets mining [2].

In this paper, we have discussed the various ARM algorithms in four computing paradigm, sequential, parallel and distribute, grid, and cloud computing. Although this study may not be complete with respect to including all the algorithms but tried to cover maximum. Initially many ARM algorithms have been developed which are based on sequential computation. To achieve more speed up and scalability, parallel and distributed computation has been applied on basic algorithms and consequently parallel and distributed ARM algorithms have been proposed. With the emergence of grid computing, grid based ARM algorithms have been designed for grid computing and heterogeneous environment. The current state-of-the-art is to design efficient algorithms in cloud computing environment. Cloud computing uses MapReduce programming paradigm for parallel processing of large datasets.

**Sequential ARM Algorithms**
Sequential ARM algorithms are generally categorized into three classes based on the techniques used to mine frequent itemsets. The first one is Apriori [4] and its variants, the second one is FP-Growth [5] and the third one is Eclat and Clique based [6]. Apriori based algorithms generate frequent itemsets with candidate itemsets generations while FP-Growth based algorithms generate frequent itemsets without generating candidate itemsets. Eclat, MaxEclat, Clique, MaxClique algorithms are based on equivalence classes & hypergraph clique clustering and lattice traversal schemes. Table 1 summarizes the various sequential ARM algorithms.

The algorithms described briefly in Table 1 can be classified as basic algorithms and their variants. Basic algorithms are Apriori, FP-Growth and Eclat. Variants algorithms have tried to improve the efficiency by applying some modifications on respective basic algorithms. Improvements over Apriori have addressed the issues like minimization of database scan, reducing database size, efficient pruning techniques and efficient data structures. Data structures are being a turning point in ARM algorithms and bases for new algorithms. FP-Growth, Tree-Projection and H-Mine are such examples. Some major data structures along with their significance in algorithms are discussed as follows.

**Table 1:** List of Sequential ARM Algorithms with Brief Descriptions

| Sl. No. | Algorithms | Characteristics of Algorithms | Proposed By |
|---|---|---|---|
| 1. | AIS (Artificial Immune System) | First ARM algorithm inspired by biological immune system. | Agrawal *et al.* (1993) [1] |
| 2. | OCD (Off-line Candidate Determination) | Improvement over AIS. Advantage increases with lower support threshold. | Mannila *et al.* (1994) [9] |
| 3. | SETM (Set-Oriented Mining) | Created for SQL computing and uses relational operations. | Houtsma *et al.* (1995) [10] |
| 4. | Apriori | Seminal algorithm for mining association rule. Most simple and well known algorithm. | Agrawal *et al.* (1994) [4] |
| 5. | Apriori-TID, Apriori-Hybrid | Apriori-TID has additional feature of not using database for support counting after first pass. Apriori-Hybrid uses Apriori in initial passes and after few passes switches to Apriori-TID. | Agrawal *et al.* (1994) [4] |
| 6. | DHP (Direct Hashing and Pruning) | Apriori based algorithm. A hash-based algorithm efficiently generates large itemsets and effectively reduces database size. | Park *et al.* (1995b) [11] |
| 7. | Partitioning | Database is divided into non-overlapping partitions. Partition size must be able to accommodate in main memory. A vertical tidlist is maintained for each item. | Savasere *et al.* (1995) [12] |
| 8. | SEAR (Sequential Efficient Association Rule), SPEAR (Sequential Partitioned Efficient Association Rule) | SEAR uses prefix tree data structure while Apriori uses hash tree. SPEAR is a SEAR with non-tidlist version of Partitioning. SEAR was also optimized using pass-bundling. | Mueller (1995) [7] |
| 9. | Sampling | Finds association rule using random sample of database and verifies the results with the remaining database. Use lower support value. | Toivonen (1996) [13] |
| 10. | DIC (Dynamic Itemset Counting) | Database is partitioned into blocks marked by start points. Candidate itemsets are added at different points during a scan. New candidate itemsets can be added at any start point, unlike in Apriori. | Brin *et al.* (1997) [14] |
| 11. | CARMA (Continuous Association Rule Mining Algorithm) | The support and confidence thresholds can be changed at any time. Can be used in transaction sequence from a network. | Hidber (1999) [15] |
| 12. | Eclat, MaxEclat, Clique, MaxClique | Eclat and MaxEclat apply bottom-up and hybrid search respectively but the same prefix-based classes. Clique and MaxClique apply bottom-up and hybrid search respectively but the same clique-based classes. | Zaki *et al.* (1997b) [6] |
| 13. | FP-Growth (Frequent Pattern Growth) | Based on FP-tree. It is an extended prefix tree that stores compressed database to avoid repeated scans. Do not generate large number of candidates, unlike Apriori. | Han *et al.* (2000) [5] |
| 14. | PASCAL | Optimizes Apriori based on pattern counting inference. It introduced a new concept called key patterns. | Bastide *et al.* (2000) [16] |
| 15. | Tree-Projection | Use transaction projection with lexicographic tree for generating frequent itemset. | Agarwal *et al.* (2001) [17] |
| 16. | H-Mine | Introduces a new data structure, H-struct and a new method of mining, H-mine. H-mine is a memory-based Hyper Structure Mining. Unlike FP-Growth, H-mine does not create any physical projected database or conditional FP-Tree. | Pei *et al.* (2001) [18] |
| 17. | RElim (Recursive Elimination) | Without prefix tree or H-struct process the transaction directly, organizing them merely into singly linked list. Focuses more on simplicity than performance although it's not slower. | Borgelt (2005) [19] |

Apriori algorithm proposed in [4] has used Hash Tree data structure to store and generate candidate itemsets and for support counting. Prefix Tree based Apriori algorithm named as SEAR has been proposed in [7]. Prefix Tree also known as Trie has been found faster, consumes less memory and simpler in operation [8] than Hash Tree. Eclat and Clique based algorithms use Vertical Data Layout and do not maintain any complicated data structures like Hash Tree [6]. The frequent pattern tree or FP-tree is an extended prefix tree that leads to an FP-tree based pattern fragment growth (FP-Growth) mining method [5]. It stores compressed database to avoid repeated scans. Tree Projection algorithm uses matrix counting on reduced set of transactions obtained by successive projection of transactions at the nodes of a Lexicographic Tree [17]. A new data structure H-struct is proposed by Pei *et al.* [18]. H-struct is a hyper-linked data structure that evolves a new frequent mining algorithm, H-mine.

Serial computation is the bottleneck for sequential ARM algorithms. Sequential algorithms running in serial computing environment cannot provide high scalability and high response time due limited storage and computational power. To overcome the limitations of sequential algorithms, high performance parallel and distributed ARM algorithms are proposed which are scalable and taking less response time.

**Parallel and Distributed ARM Algorithms**

When it comes to mine large volume of datasets, the sequential algorithms have been failed to prove scalability and efficiency since single computer has limitation on its memory and processing capacity. Hence parallel and distributed algorithms have been developed to perform large-scale computing in ARM algorithms on multiple processors. In order to achieve good performance of parallel algorithms one has to focus on the some challenges as minimization of synchronization, communication & disk I/O, proper load balancing, appropriate data layout and data decomposition [20]. M. J. Zaki in his papers [20-21], has surveyed and classified the parallel and distributed algorithms up to that date on the basis of different dimensions that include hardware architecture, type of parallelism, work load balancing strategy and database layout, database partition, candidate partition. Another survey paper by Dunham *et al.* [22] have explained the many sequential and parallel ARM algorithms in detailed and also classified and compared them but covered limited number of algorithms. We have briefly described the various parallel and distributed algorithms in Table 2. Further Tables 3-9 have categorized these parallel and distributed algorithms on the basis of key characteristics. The key characteristics include memory system (h/w architecture), parallelism, load balancing, database layout, database and candidate partition and base algorithm [20-21].

**Table 2:** List of Parallel and Distributed ARM Algorithms with Brief Descriptions

| Sl. No. | Algorithms | Characteristics of Algorithms | Proposed By |
|---|---|---|---|
| 1. | PDM (Parallel Data Mining) | It is DHP based and uses a hash table to generated candidates at early passes. Candidate generation in subsequent passes is same as in Apriori. Local candidates generated by each processor are exchanged through an all-to-all broadcast to obtain global candidates. | Park *et al.* (1995a) [25] |
| 2. | PEAR (Parallel SEAR Algorithm), PPAR (Partition Parallel Association Rules) | These are parallel version of SEAR and SPEAR. In PEAR, same candidate prefix tree is generated by each processor. Each processor count local supports and then global supports are produced on each processor. In PPAR, all sized local frequent itemsets are generated by local partition of databases on each processor. Processors broadcast potentially frequent itemsets to each other. | Mueller (1995) [7] |
| 3. | CD (Count Distribution), DD (Data Distribution), Cand.D (Candidate Distribution) | Parallelization of Apriori. CD and DD are based on the data parallelism and task parallelism respectively while Cand.D is based on hybrid of data parallelism and task parallelism. CD replicates all the candidates on all the processors, and distributes the database across all the processors. DD partitions the candidate set and distributes across the processors as is the database. | Agrawal *et al.* (1996) [26] |
| 4. | CCPD (Common Candidate Partitioned Database), PCCD (Partitioned Candidate Common Database) | Implementation of Apriori in shared-memory environments. CCPD has a common candidate hash tree, but logically splitted database across all processors. PCCD constructs a local partitioned candidate hash tree per processor, but common database across all processors. | Zaki *et al.* (1996) [27] |
| 5. | NPA (Non Partitioned Apriori), SPA (Simply Partitioned Apriori), HPA (Hash Partitioned Apriori), HPA-ELD (HPA with Extremely Large Itemset Duplication) | Techniques in these algorithms are much alike to CD, DD and Cand.D. In NPA, candidates are replicated across all the processors. Each processor counts support locally and finally global support counting is performed by a coordinator processor. SPA partitions the candidate itemsets across all the processors equally. Each processor broadcast its transactions to others. HPA distribute candidate across different processors using hash technique. It does not broadcast database partition across processors but moves the itemsets of transactions to the particular processor by hash technique. In HPA-LED, itemsets are sorted according to occurring frequency. Most frequently occurred itemsets are replicated across processors to reduce communication. It reduces the influence of skewed data. | Shintani *et al.* (1996) [28] |
| 6. | FDM (Fast Distributed Mining) | Generates smaller set of candidates by exploring relations between locally large and globally large itemsets. Two pruning techniques, local pruning and global pruning are used. Reduces the number of messages to be passed for support exchange. Three versions of FDM are proposed based on different combination pruning: FDM-LP, FDM-LUP and FDM-LPP. | Cheung *et al.* (1996) [29] |

| 7. | IDD (Intelligent Data Distribution), HD (Hybrid Distribution) | IDD is an improved version of DD. To reduce the redundant computation of DD, it partitions the candidates across the processors on the basis of same prefix. It switches to CD when total number of candidates fit in memory. Local portions of database are communicated by the linear ring based all to all broadcast. HD is a combination of CD and IDD. *P* processors are divided into *G* equal size groups, with *P/G* processors in each group. CD is applied considering one group as one processor. | Han *et al.* (1997) [30] |
| 8. | PE (ParEclat), PME (ParMaxEclat), PC (ParClique), PMC (ParMaxClique) | Based on their sequential counterparts. The four algorithms are similar with respect to parallelism technique and different with respect to search strategy and equivalence class clustering. Each algorithm contains three phases, initialization, asynchronous and reduction. In initialization phase, data partition and computation are carried out. Each processor generates frequent itemsets asynchronously in second phase. Last phase aggregates the final results [Zaki (1999)]. | Zaki *et al.* (1997c) [31] |
| 9. | FPM (Fast Parallel Mining) | It adopts CD and overcomes the problem with FDM. It reduces the number of candidates by applying distributed pruning and global pruning. In each of iteration, FMP requires only one round of message passing whereas FDM requires at least two rounds. | Cheung *et al.* (1998a) [32] |
| 10. | APM (Asynchronous Parallel Mining) | DIC based algorithm. It has lesser number of database scanning and smaller set of candidates than CD. Due to asynchronous nature it produces less I/O conflict. Also employs a Virtual Partition Pruning. | Cheung *et al.* (1998b) [33] |
| 11. | MLFPT (Multiple Local Frequent Pattern Tree) | FP-Growth based algorithm. Consists of two stages. Stage one constructs parallel frequent tree on each processor called Multiple Local Parallel Trees (MLPT) and stage two mines parallel frequent items using MLPT tree. | Zaïane *et al.* (2001) [34] |
| 12. | Parallel FP-Growth | Presents parallel execution of FP-Growth. Also examine the challenges to parallelize and method to balance the execution efficiently on shared-nothing architecture. | P. I. *et al.* (2003) [35] |

These parallel and distributed algorithms described in Table 2 are not able to address all challenging issues related to mining of huge, distributed and remote data sets as they are designed for homogeneous computing environment [23]. Most of the existing parallel and distributed ARM algorithms are based on static load balancing and partition the database evenly across the computing nodes because of the homogeneous environment. Therefore, they could not have employed well on emerging grid computing infrastructure as well as on heterogeneous compute cluster. Due to the heterogeneity of computing nodes in grid/cluster system and also inherent dynamic nature of association rule mining, execution of traditional parallel and distributed algorithms degrade the performance [24]. To overcome such issues new algorithms have been developed for grid computing environment.

**Table 3:** Classification of Parallel and Distributed ARM Algorithms based on Memory System

| Distributed | Shared | Hierarchical |
|---|---|---|
| CD, PDM, DD, IDD, NPA, SPA, HPA, HPA-ELD, Cand.D, PEAR, PPAR, HD, FDM, FPM | CCPD, PCCD, APM, MLFPT | ParEclat, ParMaxEclat, ParClique, ParMaxClique, Parallel FP-Growth |

**Table 4:** Classification of Parallel and Distributed ARM Algorithms based on Parallelism

| Data Parallelism | Task Parallelism | Hybrid |
|---|---|---|
| CD, PDM, CCPD, NPA, PEAR, FDM, FPM | PCCD, DD, IDD, SPA, HPA, HPA-ELD, Cand.D, PPAR, APM, ParEclat, ParMaxEclat, ParClique, ParMaxClique | HD |

**Table 5:** Classification of Parallel and Distributed ARM Algorithms based on Load Balancing

| Static | Hybrid |
|---|---|
| CD, PDM, CCPD, PCCD, DD, IDD, NPA, SPA, HPA, HPA-ELD, Cand.D, PEAR, PPAR, FDM, FPM, APM, ParEclat, ParMaxEclat, ParClique, ParMaxClique | HD, MLFPT, Parallel FP-Growth |

**Table 6:** Classification of Parallel and Distributed ARM Algorithms based on Database Layout

| Horizontal | Vertical |
|---|---|
| CD, PDM, CCPD, PCCD, DD, IDD, NPA, SPA, HPA, HAP-ELD, Cand.D, PEAR, PPAR, HD, FDM, FPM, APM, MLFPT, Parallel FP-Growth | ParEclat, ParMaxEclat, ParClique, ParMaxClique |

**Table 7:** Classification of Parallel and Distributed ARM Algorithms based on Database Partition

| Partitioned | Replicated/Partially Replicated | Shared |
|---|---|---|
| CD, PDM, CCPD, DD, IDD, NPA, SPA, PEAR, PPAR, HD, FDM, FPM, APM | HPA, HAP-ELD, Cand.D, ParEclat, ParMaxEclat, ParClique, ParMaxClique | PCCD |

**Table 8:** Classification of Parallel and Distributed ARM Algorithms based on Candidate Partition

| Partitioned | Replicated | Shared | Hybrid |
|---|---|---|---|
| PCCD, DD, IDD, SPA, HPA, HPA-ELD, Cand.D, ParEclat, ParMaxEclat, ParClique, ParMaxClique | CD, PDM, NPA, PEAR, PPAR, FDM, FPM | CCPD, APM | HD |

**Table 9:** Classification of Parallel and Distributed ARM Algorithms based on Base Algorithms

| Apriori | Partition | DIC | Eclat, Clique | FP-Growth |
|---|---|---|---|---|
| CD, PDM, CCPD, PCCD, DD, IDD, NPA, SPA, NPA, HPA-ELD, CandD, PEAR, HD, FDM, FPM | PPAR | APM | ParEclat, ParMaxEclat, ParClique, ParMaxClique | MLFPT, Parallel FP-Growth |

**ARM Algorithms in Grid Computing Environment**

Parallel association rule mining deals with tightly coupled system: shared or distributed memory machine and fast network enabled clusters. Distributed association rule mining deals with loosely-coupled system like cluster in which nodes are connected by slow networks and geographically distributed. Grid association rule mining is designed to support data distribution on geographically distributed nodes and use computing resources of these nodes in order to achieve high performance [24]. Grid computing, an emerging technology provides a global computing infrastructure by integrating the heterogeneous computing system and data sets. There are following reasons which demands computational grid in data mining [23]. (1) There is need to access multiple databases and data sources because a single database does not able to hold all required data in an application. (2) Multiple databases may not always belong to the same organization and may not be found at the same location due to the geographical distribution. (3) To enhance the performance of data mining process, local copies of the whole datasets or subset of it may be used.

Research in data mining on grid computing environment can be categorized into two directions, grid based data mining applications and grid based data mining algorithms. Both the approaches are described in the following sub-sections.

**Grid Based Data Mining Applications**

A number of systems have been proposed which are different from traditional parallel and distributed data mining system and are operated on clusters of computers, or over the internet. Some well known high-performance and distributed mining systems, which make use of JAVA and specially designed over cluster and internet, are listed in Table 10.

Besides these distributed applications for data mining there are so many research organizations who have worked on exploiting data mining on computational grids. Many research projects and approaches have tried to integrate and deploy data mining and grid infrastructure. Table 11 lists them and describes their different approaches to support Data Mining over Distributed or Grid Infrastructure.

**Table 10:** Distributed Mining Systems designed over cluster and internet using Java

| Sl. No. | Systems/Applications/Projects | Developed/Proposed By |
|---|---|---|
| 1. | Java Agent for Meta Learning (JAM) | Stolfo *et al.* (1997) [36] |
| 2. | Kensington Enterprise Data Mining System | Chattratichat *et al.* (1999) [37] |
| 3. | Besizing KnOwledge through Distributed Heterogeneous Induction (BODHI) | Kargupta *et al.* (1999) [38] |
| 4. | Papyrus | Grossman *et al.* (1998) [39] |
| 5. | Parallel and Distributed Data Mining Application Suit (PaDDMAS) | Rana *et al.* (2000) [40] |

Various grid data mining projects/applications described in Tables 10 and 11, only integrate and deploy classical algorithms/tools on grid and do not provide any efficient mechanism for data partitioning & distribution, load balancing and minimization of synchronization & communication. So efficient algorithms suitable for grid architecture have needed to be designed, which are based on appropriate data partitioning & distribution approach, load balancing strategy, optimization of synchronization and communication.

**Table 11:** Data Mining Applications on Grid Infrastructure

| Sl. No. | Projects/Applications | Developed/Proposed By |
|---|---|---|
| 1. | TeraGrid | Catlett (2002) [41] |
| 2. | Datacentric Grid | Skillicorn *et al.* (2002) [42] |
| 3. | Discovery Net (D-NET) | ur in *et al.* (2002) [43] |
| 4. | InfoGrid | Giannadakis *et al.* (2003) [44] |
| 5. | GridMiner | Brezany *et al.* (2003), Brezany *et al.* (2004), Hofer *et al.* (2004), Kickinger *et al.* (2004) [45-48] |
| 6. | Grid-based AdapTive Execution on Streams (GATES) | Agrawal *et al.* (2003), Chen *et al.* (2004) [49-50] |
| 7. | The Knowledge Grid | Cannataro *et al.* (2003) [51] |
| 8. | WekaG | Pérez *et al.* (2005), Pérez *et al.* (2007) [23] [52] |
| 9. | Weka4WS | Talia *et al.* (2005) [53] |
| 10. | Grid Weka | Khoussainov *et al.* (2004) [54] |
| 11. | DataMiningGrid | Stankovski *et al.* (2008a), Stankovski *et al.* (2008b) [55-56] |
| 12. | Federated Analysis Environment for Heterogeneous Intelligent Mining (FAEHIM) | Shaikh *et al.* (2008) [57] |

**Grid Based ARM Algorithms**

The non homogeneity of grid leads to unbalanced distribution of datasets and workloads. The distributed frequent itemsets generation is a very computation intensive task, and a lack of proper load balancing and synchronization technique will lead to significant degradation in performance [58]. ARM is an iterative and interactive process and at the end of each of iteration, a phase of synchronization is required in order to count global support. If the database is distributed randomly across the processors then processors will not complete their job at the same time, which increases idle time of processors and finally increasing the overall execution time of the system. The balanced distribution of the database among processors does not guarantee that processors will reach synchronization barrier at the same time. The reason behind this is that degree of correlation between items in each partition does not necessarily same. Frequent itemsets mining exhibits dynamic nature due to its degree of correlation between itemsets which is unpredictable before execution. Thus there is a need of dynamic load balancing strategy during execution under grid environment. So in order to minimize the load imbalance in grid based ARM algorithm one have to design an appropriate data partitioning approach before execution and a dynamic load balancing strategy during execution [59]. Various ARM algorithms for grid environment have been proposed, which focus on such issues. Table 12 lists some algorithms with their techniques of data distribution and load balancing.

**Table 12:** ARM Algorithms on Grid Computing Infrastructure

| Sl. No. | Algorithms | Characteristics | Proposed By |
|---|---|---|---|
| 1. | DisDaMin (DIStributed DAta MINing) Project | Uses Intelligent Data Distribution based on clustering methods. | Fiolet *et al.* (2006) [60] |
| 2. | GridDMM (Grid-based Distributed Max-Miner) | Evenly distributes database across nodes. | Luo *et al.* (2006) [61] |

| 3. | Distributed Algorithm for Frequent Itemsets Generation on Heterogeneous Clusters and Grid Environments | Focuses on inherent dynamic nature of ARM and memory constraints of nodes. Uses dynamic load balancing using block-based partitioning. | Aouad *et al.* (2007) [58] |
|---|---|---|---|
| 4. | HDDS (Heuristic Data Distribution Scheme) | Formulates the data distribution problem by linear programming and heuristic to solve it. | Yang *et al.* (2008) [62] |
| 5. | Hybrid Load Balancing Strategy | Proposed a novel data partitioning approach, a hierarchical model of grid and dynamic load balancing strategy. | Tlili *et al.* (2011b), Tlili *et al.* (2011a), Tlili *et al.* (2012b), Tlili *et al.* (2012a), Tlili *et al.* (2012c) [24] [59] [63-65] |

**ARM Algorithms in Cloud Computing Environment**
Cloud computing delivers hardware and software as a service over the internet in a scalable and simple way. Cloud computing provides efficient platform for storage and analysis of huge data [66]. Data mining through cloud computing benefits the small business as they need not to own the software and hardware since these are available on reduced cost by cloud service providers [67]. There are many companies and organizations, which provide cloud services. Amazon's web services Elastic Compute Cloud (EC2) [http://aws.amazon.com/ec2] and Simple Storage Service (S3) [http://aws.amazon.com/s3] provide computational services on demand and large space for storing & accessing data in a secure manner. Microsoft also launched its cloud based Azure Services Platform [http://windowsazure.com/en-us/] similar to Amazon, facilitates environments in which anyone can develop, host and manage cloud based applications. Google have introduced a new technique to store and process excessive data. The hardware infrastructure of Google is based on the scalable clusters consisting of large number of commodity computers instead of using high performance computers. It has introduces Google File system (GFS) [68], a new storage system and MapReduce framework [69], a new computing paradigm.

MapReduce introduces a paradigm shift in parallel processing. It has overcome the customary overheads of the traditional distributed framework. In traditional distributed system, one has to take care of node failures that lead to re-executions of the jobs. Message Passing Interface (MPI) programming paradigm used in scientific distributed system requires taking care of data partition, synchronization, communication and scheduling [70-71]. In MapReduce computing power is moved to where the data is rather than moving data as in MPI. MapReduce is a simplified programming paradigm for distributed processing of large scale data. Cloud systems providing data processing and analytics use MapReduce framework. Hadoop [http://hadoop.apache.org] is an open source project of Apache [http://www.apache.org] that has implemented Google's cloud approach. It provides Hadoop Distributed File System (HDFS) [72] as data storage system based on GFS and Hadoop MapReduce [73] as data processing system based on Google's MapReduce. Now Hadoop is a large-scale distributed batch processing infrastructure for parallel processing of big data on large cluster of commodity computers. Many data mining algorithms including ARM algorithms have been re-designed on MapReduce framework for parallel processing of large scale data.

A data processing application runs as a MapReduce job on local/cloud Hadoop cluster. The datasets to be processed must be resided in HDFS. A MapReduce job consists of Mapper, Combiner and Reducer tasks. A MapReduce job is splitted into multiple tasks and all the tasks run concurrently on the available nodes in the cluster. The input and output of Mapper, Combiner and Reducers must be in form of *(key, value)* pairs. More about the MapReduce programming paradigm and HDFS can be found in [73]. Several ARM algorithms have been proposed on MapReduce framework and most of them are Apriori based. Table 13 lists various MapReduce based ARM algorithms with a brief description along with corresponding paper's title.

**Table 13:** MapReduce Based ARM Algorithms

| Sl. No. | Paper's Title | Algorithm and its Description | Proposed By |
|---|---|---|---|
| 1. | MapReduce as a programming model for association rules algorithm on Hadoop | A straight forward implementation of Apriori on MapReduce. | Yang *et al.* (2010) [74] |
| 2. | The Strategy of Mining Association Rule Based on Cloud Computing | Proposed a one phase MapReduce implementation of Apriori and a data distribution method that is applied on it. | Li *et al.* (2011) [66] |
| 3. | An improved Apriori Algorithm Based on the Boolean Matrix and Hadoop | Proposed an improved Apriori on Hadoop. Transactional database is converted to Boolean matrix and operated by AND operation. | Yu *et al.* (2011) [75] |
| 4. | Data Mining Using Clouds: An Experimental Implementation of Apriori over MapReduce | A straight forward implementation of Apriori on MapReduce. | Li *et al.* (2012a) [76] |
| 5. | Apriori-based frequent itemset mining algorithms on MapReduce | Proposed three Apriori based Algorithms on MapReduce, SPC, FPC and DPC. SPC is straight forward implementation. FPC and DPC are based on combined passes in a single MapReduce phase. | Lin *et al.* (2012) [70] |
| 6. | Parallel Implementation of Apriori Algorithm based on MapReduce | A straight forward implementation of Apriori on MapReduce. | Li *et al.* (2012b) [77] |
| 7. | PARMA: A Parallel Randomized Algorithm for Approximate Association Rules Mining in MapReduce | PARMA is implemented on MapReduce. Discovers approximate frequent itemsets and association rules by mining a small random sample of the datasets. | Riondato *et al.* (2012) [78] |
| 8. | Frequent Itemset Mining on Hadoop | Straight forward implementation of Apriori on MapReduce. Generate candidates inside reducer instead of mapper. 1 and 2 –itemsets are counted in a single step using triangular matrix. | Kovacs *et al.* (2013) [79] |
| 9. | Exploring HADOOP as a Platform for Distributed Association Rule Mining | A straight forward implementation of Apriori on MapReduce. | Oruganti *et al.* (2013) [80] |
| 10. | MR-Apriori: Association Rules Algorithm Based on MapReduce | MR-Apriori algorithm is a straight forward implementation of Apriori on MapReduce. | Lin (2014) [81] |
| 11. | PFP: Parallel FP-Growth for Query Recommendation | PFP algorithm is a MapReduce implementation of parallel FP-Growth algorithm. | Li *et al.* (2008) [82] |
| 12. | Balanced Parallel FP-Growth with MapReduce | BPFP algorithm. Improves parallelization of PFP by applying load balance feature. | Zhou *et al.* (2010) [83] |
| 13. | A parallel Algorithm of Association Rules Based on Cloud Computing | Proposed a cloud based PFP-growth algorithm using MapReduce model. | Yong *et al.* (2013) [84] |
| 14. | Frequent Itemset Mining for Big Data | Proposed two algorithms, Dist-Eclat and BigFIM. Dist-Eclat is based on Eclat while BigFIM first uses Apriori based method and later on switches to Eclat. | Moens *et al.* (2013) [71] |
| 15. | Performance Analysis of Apriori Algorithm with Different Data Structures on Hadoop Cluster | Apriori with data structure hash table trie has been found best performing in MapReduce context while it was not efficient for sequential Apriori. | Singh *et al.* (2015) [85] |

Apart from the above proposed algorithms on MapReduce framework, Apache provides tools running on Hadoop that implemented many data mining and machine learning algorithms. Apache's Mahout [https://mahout.apache.org] is a library of various algorithms of machine learning and data mining. Apache's Spark [http://spark.apache.org] also provides machine learning library *spark.mllib* [86].

## Conclusion

This paper surveyed the ARM algorithms in various computing paradigm. The considered paradigms are sequential computing, parallel and distributed computing, grid computing, and cloud computing. The family of sequential algorithm is classified into three branches, Apriori based, Eclat and Clique based, and FP-Growth based. Parallel and Distributed algorithms have been designed to increase scalability and decrease the execution time that was not possible with sequential computation. These parallel and distributed algorithms have been proposed for homogeneous cluster systems shows degradation in the performance when applied on heterogeneous cluster or gird infrastructure. The inherent dynamic nature of association rule mining and heterogeneity of computing nodes in grid/cluster are the major causes that degrade the performance of existing algorithms. This leads to the design of new algorithms for grid infrastructure. Initially grid based data mining applications have been developed that integrates traditional data mining techniques with grid infrastructure. To better utilize the grid resources and enhance the performance, efficient algorithms suitable for grid architecture have been designed, which are based on appropriate data partitioning and distribution, load balancing strategy, optimization of synchronization and communication. Hadoop MapReduce shifts the paradigm of parallel and distributed computing. It moves the computing power to where the data is rather than moving data. It is one of the most popular infrastructures to process and analyze big data. Many ARM algorithms have been re-designed on MapReduce framework for processing large scale data. Data mining and machine learning libraries are also developed to be run on Hadoop cluster.

October 2015.